\documentclass[prb,showpacs,floatfix,twocolumn]{revtex4}
\usepackage{graphicx}
\usepackage{amssymb}
\usepackage{dcolumn}
\usepackage{bm}

\usepackage{color}

\begin{document}

\title{Aftershock production rate of driven viscoelastic interfaces}
\author{E. A. Jagla}

\affiliation{Centro At\'omico Bariloche and Instituto Balseiro, Comisi\'on Nacional de Energ\'{\i}a At\'omica, 
(8400) Bariloche, Argentina}

\begin{abstract}

We study analytically and by numerical simulations the statistics of the aftershocks generated after large avalanches in models
of interface depinning that include viscoelastic relaxation effects. We find in all the analyzed cases that the decay law of aftershocks with time
can be understood by considering the typical roughness of the interface and its evolution due to relaxation. In models where there is a single 
viscoelastic relaxation time there is an exponential decay of the number of aftershocks with time. In models in which viscoelastic relaxation is wave-vector dependent we typically find a power law dependence of the decay rate, compatible with the Omori law.
The factors that determine the value of the decay exponent are analyzed.

\end{abstract}
\maketitle

\section {Introduction}

An elastic interface driven through a disordered energy landscape is a generic model for
many different physical systems, as domain walls in ferromagnetic materials \cite{ferro1,ferro2,ferro3}, wetting fronts on a rough
substrate \cite{wet1,wet2}, and seismic fault dynamics \cite{eq1,eq2,eq3}.
The characteristic feature of the dynamics of slowly driven elastic interfaces is its evolution through a sequence of abrupt events, called avalanches.
In the presence of viscoelastic effects, interface depinning has additional interesting physical properties \cite{marchetti,landes,papanicolau}, one of them being the existence of aftershocks, namely secondary avalanches originated in the internal viscoelastic dynamics, that are not directly related to the external driving. The prominent example of a physical system in which aftershocks show up is the seismic phenomenon\cite{scholz}. There, aftershocks are so abundant (they may even represent the   numerical majority of the events) that strong statistical regularities are well established for them since many years ago. The most famous of these empirical observations is the Omori (or Omori-Utsu) law \cite{omori,utsu1,utsu2}, stating that the aftershock rate (the number of aftershocks per unit of time) $N(t)$ after a main shock decays as

\begin{equation}
N(t)\sim \frac 1{(t+c)^p}.
\label{omori}
\end{equation}
Here time is measured from the time of the main shock, and $p$, $c$ are phenomenological parameters. The value of $c$ is typically in the range of minutes. The most interesting physical information contained in the Omori law is the fact that for $t\gg c$, $N(t)$ decays as a power law with an exponent $p$. Experimentally, although typically a value of $p$ around 1 is referred to, a much wider range of values (between 0.9 and 1.5 according to \cite{utsu2}) has been observed. Also it has to be taken into account that the fitting of experimental data with the power law (\ref{omori})
has usually important deviations. In spite of this, the fact that the aftershock rate is roughly a power law with time is well established. 

There have been different proposals for the physical origin of aftershocks. Some studies \cite{perfett} have claimed that they are related to aseismic afterslip occurring after large quakes. Evidence for this mechanism is not compelling. In any case, the model we will study does not include the possibility of afterslip, meaning that the aftershocks we will observe are not related to this physical mechanism.

The most accepted theory of aftershock production follows the analysis in \cite{dietrich1994}. After a main shock, some parts of the fault close to the ruptured region are suddenly loaded to higher values of stress. This can produce the failure of these regions in a finite time, according to the mechanisms of static fatigue rupture. Within this framework, the Omori law has been derived by a number of authors \cite{dietrich1994,marcellini,helmstetter,zoller}. Note that according to this mechanism, aftershocks should mostly appear outside and nearby the region affected by the main shock.

In the last years, it has been shown that the statistical properties of earthquakes in single fault systems can be well described by the avalanches observed in viscoelastic models of interface depinning \cite{landes,jagla,kolton}. Numerical simulations have shown that these kind of models are able to reproduce many statistical properties of earthquakes, like the Gutenberg-Richter law with a realistic $b$ exponent, and also friction properties of the system compatible with experimental observations and with the predictions of the phenomenological rate-and-state equations\cite{scholz}.
In addition, these models display aftershocks that qualitatively resemble real ones. 

The purpose of the present work is to study in detail the aftershock rate in these kind of models, and see whether this rate is compatible with the Omori law or not.

\section {The model}

\begin{figure}[h]
\includegraphics[width=6cm,clip=true]{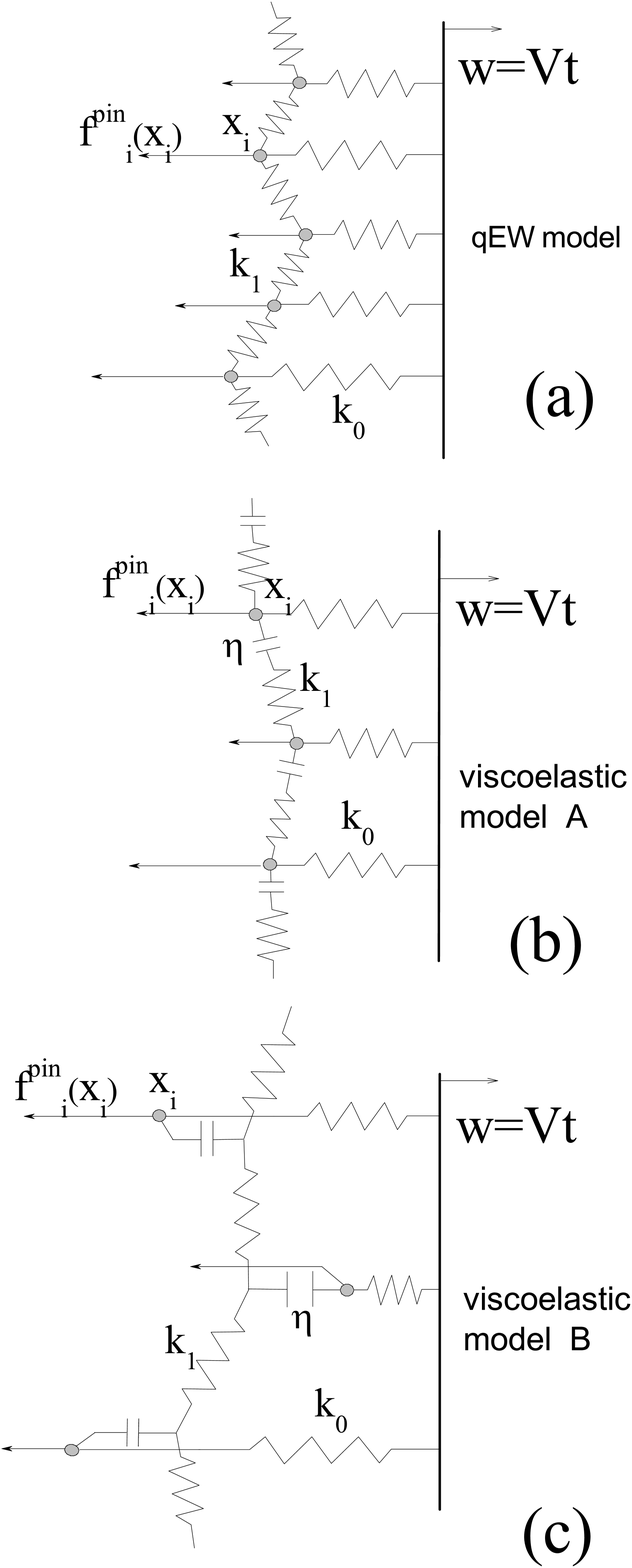}
\caption{(a) Mechanical representation of the quenched Edwards-Wilkinson model. (b-c) Two viscoelastic variations of the model.
}
\label{f1}
\end{figure}

The models we study here are based on the standard quenched Edwards-Wilkinson (qEW) model that describes the dynamics of a purely elastic interface on a disordered substrate \cite{eq3}. 
A schematic pictorial view of this model is presented in Fig. \ref{f1}(a). For simplicity the sketch is made  for the one-dimensional case, but we will discuss the two-dimensional case throughout the paper, which is the appropriate case for the seismic context. The dynamical state of the model is described by the coordinates $x_i$ at every spatial position $i$. It is convenient to define $f_i$ as the total elastic force exerted over the site $i$, except the force exerted by the $k_0$ springs. In the present case this is given by
\begin{equation}
f_i=k_1(\nabla^2 x)_i.
\label{fi}
\end{equation}
Here, $(\nabla^2 x)_i\equiv \sum_j(x_j-x_i)$ ($j$ being the neighbor sites to $i$) is the discrete Laplacian operator.

The driving velocity $V$ is supposed to be vanishingly small compared with the dynamics of the $x_i$ variables. This means that avalanches are instantaneous in the time scale of driving,  a condition that is quite well satisfied in actual seismicity.
For numerical convenience we consider a case in which the substrate potential is a collection of discrete narrow wells at which the interface (through the variables $x_i$) can be trapped. Each well is characterized by the maximum force $f^{th}_i$ that it can apply on the surface. The actual pinning force $f^{pin}_i$ cannot overpass this maximum value.
In mechanical equilibrium, the force on each $x_i$ must balance, and in our narrow well limit this means
\begin{equation}
k_0(Vt-x_i)+f_i=f^{pin}_i<f^{th}_i.
\label{equi}
\end{equation}

The dynamics of the model in the narrow well approximation can be stated as a set of rules for the evolution of the forces $f_i$ acting on each site $i$
\cite{landes}. 
On a stationary configuration satisfying (\ref{equi}), time increases until $k_0(Vt-x_i)+f_i=f^{th}_i$ for some $i$. At this stage the corresponding $x_i$ jumps to the next potential well located at $x_i+z$ ($z$ is taken from a random distribution, defined in the Appendix). Due to this rearrangement, and according to (\ref{fi}), the forces $f$ are modified as 
\begin{eqnarray}
f_i&\leftarrow&f_i-4k_1z\nonumber\\
f_j&\leftarrow&f_j+k_1z
\label{avalanche1}
\end{eqnarray}
where $j$ are the four sites (in two dimensions) neighbor to $i$.
This can generate a cascade of rearrangements that represents an earthquake in the model. The avalanche finishes when the stability condition (\ref{equi}) holds again at every site.

In general terms, viscoelastic relaxation is a mechanism by which the mechanical energy of the system tends to be reduced in time, taking the system to more relaxed configurations. The existence of these mechanisms is well documented in the earthquake context, and have manifestations at the laboratory scale, where for instance they are responsible for the slow increase in time of the real contact area between two solid bodies at rest\cite{kilgore}. 
Within the context of our numerical models, these mechanisms can be conveniently represented by additional terms in the time evolution equation of the model. In some cases, they can be graphically represented by means of linear viscoelastic elements.

It is not obvious {\em a priori} what is the exact form of the viscoelastic terms that must be used to model seismic processes in the most accurate way. It is for this reason that we consider two different forms of the relaxation mechanism that display different properties for the aftershock activity. 
They are graphically represented in Fig. \ref{f1}(b-c). 
The viscous elements are all identical and characterized by a viscosity coefficient $\eta$. 
The main difference between the two models is that the one in Fig. \ref{f1}(b) has a single time constant for relaxation, whereas that in Fig. \ref{f1}(c) has a distribution of relaxation times depending on wave length.

Similarly to the case of the qEW model, we define $f_i$ as the total (visco-)elastic force  exerted on the site $i$, except the force applied by $k_0$ springs, so that Eq. (\ref{equi}) is the stability condition for the viscoelastic versions also.
However, even if the values of $x_i$ remain fix (i.e., as long as $f_i^{pin}<f_i^{th}$ for all $i$), $f_i$'s are no longer constant, but evolve in time according to\cite{footnote}

\begin{eqnarray}
\frac {df_i}{dt}&=&-\frac{k_1}{\eta} f_i ~~~~~~~~~\mbox{(model A)}\label{d+r_1}\\
\frac {df_i}{dt}&=&\frac{k_1}{\eta}(\nabla^2  f  )_i~~~~~~\mbox{(model B)}\label{d+r_2}
\end{eqnarray}

As in the qEW model, an avalanche occurs each time $k_0(Vt-x_i)+f_i=f^{th}_i$ for some $i$.
The time scale of relaxation $\eta/k_1$ is supposed to be very large compared with the time scale of the individual avalanches, namely we continue to consider avalanches as instantaneous. Therefore, equations (\ref{avalanche1}) continue to be valid in the viscoelastic case, as dashpots are rigid during the avalanche development.

\begin{figure}[h]
\includegraphics[width=8cm,clip=true]{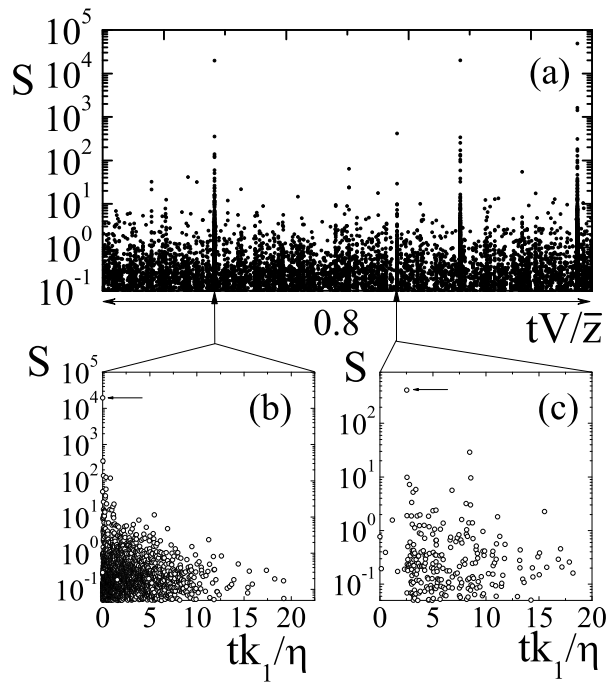}
\caption{(a) The size $S$ of the avalanches (see definition in Appendix) as a function of time, for the case of model A (Fig. \ref{f1}(b)). The time axis is scaled by the driving velocity $V$, so all aftershocks appear on the same vertical line. In (b) and (c), we take particular clusters of events in (a) and plot them as a function of the time, in units of the relaxation time constant $\eta/k_1$. We see here the decay of the aftershock rate with time. The largest event in each cluster is indicated by the small horizontal arrows. Note that in (b) the largest event is the first one, but this is not the case in (c). This last case is actually the typical situation. 
}
\label{f3}
\end{figure}

Since $f_i$ depend on time even if $x_i$ are constant, the triggering of an avalanche is now a combined effect of the driving and the viscoelastic dynamics of the system, respectively represented by the two terms on the l.h.s. of (\ref{equi}). 
Only in the case in which $V\ll \overline z k_1/\eta$ (where $\overline z$ is the average separation between potential wells) we have a complete separation of time scales, and we can tell which term is the immediate responsible for each triggered event. 
We will consider this limiting case from now on. 
In these conditions, the avalanches are clustered in time. Each cluster is initiated by the driving term. All the remaining events within the cluster are aftershocks, that are triggered by the relaxation term. Note that the first event is not necessarily the largest one in the cluster. An example of a time sequence of events obtained by numerical simulations of model A is shown in Fig. \ref{f3}. In real seismicity the time scales of driving and relaxation are not totally separated, and a clear cut identification of aftershocks is not possible.
Although it is not our main concern here, it is necessary to mention that both models A and B produce a Gutenberg-Richter distribution of number of earthquakes as a function of size, with a realistic value of the $b$ exponent.

\begin{figure}[h]
\includegraphics[width=8cm,clip=true]{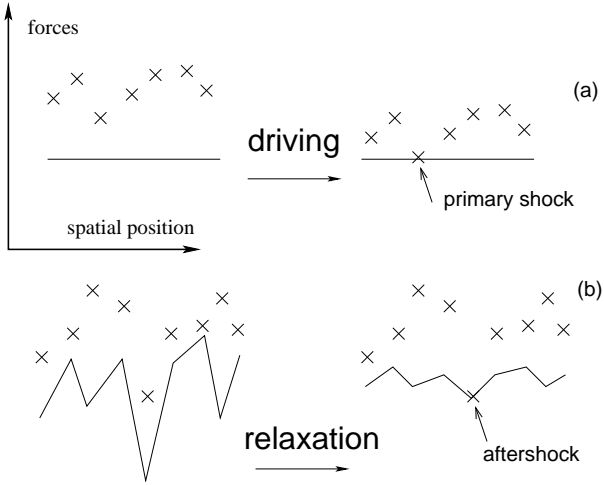}
\caption{(a) A relaxed distribution of forces $f_i$ (line), which for the present models is a constant, and the values of $f_i^{th}-k_0(Vt-x_i)$  (crosses) in a one-dimensional sketch. As driving increases, crosses move down, generating a primary shock. (b) After the primary shock, the rearranged distribution of forces $f_i$ evolves due to the relaxation term, eventually triggering aftershocks. 
}
\label{f2}
\end{figure}

We can make the following pictorial description of how initial shocks and aftershocks are produced. In  the left part of Fig. \ref{f2} (a) the form of $f_i$ is supposed to be totally relaxed, and for the models we are analyzing this means $f_i=0$ \cite{footnote}.  When, due to driving, the stability condition (\ref{equi}) is no longer satisfied at some point (right), an initial shock occurs. This produces rearrangements in the values of $f_i$ due to the avalanche dynamics (Eqs. (\ref{avalanche1})), generating a non-relaxed configuration that evolves according to (\ref{d+r_1}) or (\ref{d+r_2}). As indicated in Fig. \ref{f2}(b), this relaxation can eventually produce aftershocks at some nearby position.

This means that the number of aftershocks per unit of time are the number of sites at which $f_i$ becomes larger than $f_i^{th}-k_0(Vt-x_i)$ due to the relaxation. However, the calculation of this number is difficult because $f_i$ themselves are changed after each aftershock. In order to get a rough estimation, 
we consider a typical distribution $f_i(t=0)$ of the forces after some initial shock, and consider its evolution through relaxation, disregarding the changes of $f_i$ when aftershocks actually occur.
Assuming that the possible values of $f_i^{th}-k_0(Vt-x_i)$ are uniformly distributed, the probability to trigger an aftershock at site $i$, 
will be proportional to 
the increase of $f_i$ above all previous values it has taken before. Then,
defining $f_i^{max}(t)$ as the maximum value of $f_i$ for all times smaller than $t$ (namely $f_i^{max}(t)=\max_{0<\tau<t} f_i(\tau)$) we will estimate
\begin{equation}
N(t)\sim \sum_i \frac{d f_i^{max}}{dt}
\label{max}
\end{equation}
The effect of aftershock interaction, that we neglected in this estimation will be reconsidered later on. By now we concentrate in making an estimate of Eq. (\ref{max}).

The initial distribution of forces $f_i(0)$ needed for this estimation must be taken as a typical distribution of forces after a big shock in the system.
For the models in Fig. \ref{f1} (b-c), if the shock produced some given displacements of the surface $\delta x_i$, the values of $f_i(0)$ are given by  $f_i(0)=k_1 (\nabla^2 \delta x)_i$. This is because dashpots are rigid in the time scale of the avalanches.
Typically, statistical properties of $\delta x_i$ are characterized by a number $\zeta$ which is the roughness exponent of the displacements during the shock. It indicates that for two points on the surface a spatial distance $L$ apart, the shock produced displacements that scale with $L$ as  $|\delta x_0-\delta x_L|\sim L^\zeta$. It can also be shown that this implies a spectral
form of $\delta x$ of the form 

\begin{eqnarray}
|\delta x_q|^2&\sim& q^{-(d+2\zeta)}\\ 
|f_q(0)|^2=q^4|\delta x_q|^2&\sim& q^{4-d-2\zeta}
\label{fdeq}
\end{eqnarray}
where $d$ is the dimensionality of the surface (two in our case). 

Once we have the force distribution at the initial time, we must follow the evolution caused by relaxation. 
 In the simplest case of exponential relaxation, we can readily write $f_i(t)=f_i(0)e^{-k_1 t/\eta}$, where we see that for each $i$, $f_i(t)$ moves monotonously in time (either increasing or decreasing) to the relaxed value $f_i(t\to \infty)=0$. In this case $f_i^{max}(t)$ is either $f_i(t)$ if $f_i$ is negative, or $f_i(0)$ if $f_i$ is positive, and $N(t)$ can be calculated as 
\begin{equation}
N(t) \sim \frac d{dt}{\sum_i}'  f_i(t)
\label{aa}
\end{equation}
\begin{equation}
N(t)\sim\frac d{dt}{\sum_i}'f_i(0)e^{-k_1 t/\eta}
\label{ndet1}
\end{equation}
where the prime in the sum means that it must be extended only to the points for which $f_i$ is negative. In the end, this expression clearly shows that 
\begin{equation}
N(t)\sim \exp{(-k_1 t/\eta)}~~~~~~\mbox{(model A)}
\label{omori-expo}
\end{equation}
namely, in case of a single relaxation time, the aftershock rate decays exponentially in time with the same time constant.

For the $q$-dependent relaxation case (model B) the evolution of $f_i(t)$ does not need to be monotonous in time, which complicates the analysis. Yet, to try to make an estimate let us assume $f_i(t)$ is
indeed monotonous. If this is the case Eq.  (\ref{aa}) can still be used. Assuming also symmetry between regions at which $f_i>0$ and $f_i<0$ , Eq.  (\ref{aa}) can be re-written up to a factor of two as 
\begin{equation}
N(t)\sim \frac d{dt}\sum_i  |f_i(t)|
\label{ndet2}
\end{equation}
where now the sum is unrestricted.
To estimate $f_i(t)$ we first solve Eq. (\ref{d+r_2}) in Fourier space as:
\begin{equation}
f_q(t)=f_q(0) e^{-q^2 k_1t/\eta},
\label{fqdet}
\end{equation}
and now we write
\begin{eqnarray}
f_i(t)&\sim&\int d^2 q f_q(t) \exp(iqt)=\\
&=&\int d^2 q f_q(0)\exp(-q^2t) \exp(iqt).
\label{fidet}
\end{eqnarray}
Assuming random phases between different $f_q(0)$'s, and using (\ref{fdeq}) with $d=2$  we can estimate
\begin{eqnarray}
f_i(t)&\sim&\left( \int d^2 q |f_q(0)|^2\exp(-2q^2t)  \right )^{1/2}\\
&\sim&\left( \int d^2 q q^{2-2\zeta}\exp(-2q^2t)  \right )^{1/2}\label{cutoff}\\
&\sim& t^{-(2-\zeta)/2}
\label{fidet2}
\end{eqnarray}
From here and (\ref{ndet2}) we finally get
\begin{equation}
N(t)\sim t^{-(4-\zeta)/2}~~~~~~\mbox{(model B)}
\label{omori-q2}
\end{equation}

The first main outcome of this analysis is that contrary to the case of exponential relaxation, a $q$-dependent relaxation is naturally able to give a power law decay of the aftershock rate, compatible with the Omori law. 

In view of the approximations made in the previous derivation, we will first of all present some numerical tests 
of the accuracy of Eqs. (\ref{omori-expo}) and (\ref{omori-q2}) as estimations of the aftershock rate as given by Eq. (\ref{max}).
In order to do this we start with well characterized initial distributions of $f_i$. We take three different initial force distributions: in one case we take random values of $\delta x$, and calculate $f$
as $f_i=(\nabla^2 \delta x)_i$. This choice corresponds to a roughness exponent $\zeta=-1$. In the second case we take $f_i$ as random uncorrelated values, which corresponds to $\zeta=1$.
In the third case we use  as a starting configuration an interface generated by a simulation in the purely elastic qEW model (Fig. \ref{f1}(a)). For this kind of surface, a value $\zeta\simeq 0.75$ is well established. 

\begin{figure}[h]
\includegraphics[width=8cm,clip=true]{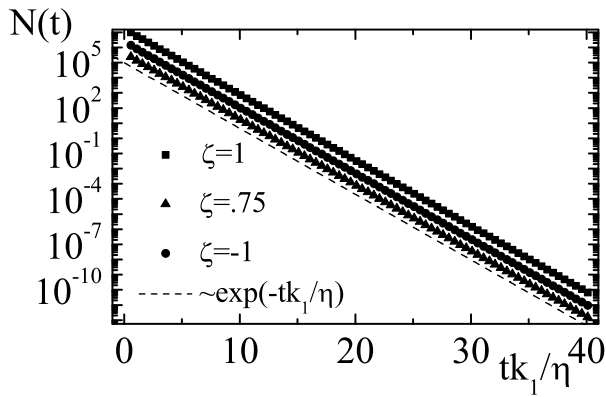}
\caption{Model A. Density of aftershocks as a function of time after the main shock, calculated according to Eq. (\ref{max})
for three different initial interfaces with different roughness (as indicated), and the corresponding analytical estimate Eq. (\ref{omori-expo}). We see the perfect accordance between the two. 
}
\label{f4}
\end{figure}

\begin{figure}[h]
\includegraphics[width=8cm,clip=true]{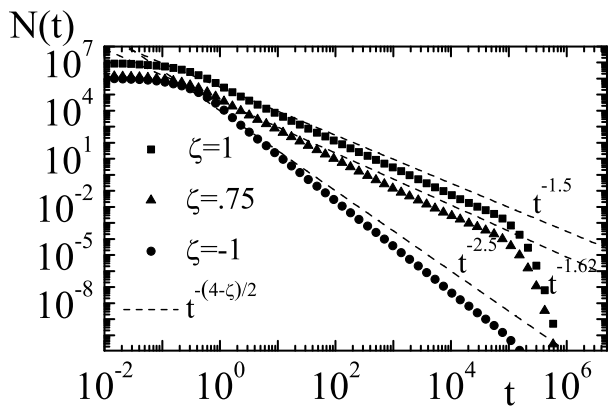}
\caption{Same as previous figure for model B, and the corresponding analytical estimate Eq. (\ref{omori-q2}). We see that the analytical estimate gives a slightly smaller decay exponent than the one calculated using Eq. (\ref{max}).
}
\label{f5}
\end{figure}

We 
evolve the initial force distribution in time, according to the appropriate relaxation mechanism (\ref{d+r_1}) or (\ref{d+r_2}). The estimated number of aftershocks is calculated using expression (\ref{max}). 
The results are presented in Figs. \ref{f4} and \ref{f5} for models A and B, and compared with the respective analytical estimates (\ref{omori-expo}) and (\ref{omori-q2}). 

For the exponential relaxation (Fig. \ref{f4}) we see that the aftershock rate calculated using Eq. (\ref{max}) is perfectly compatible with the analytical estimation Eq. (\ref{omori-expo}). This is not surprising as for this case there is no approximation in passing from Eq. (\ref{max}) to the analytical estimate Eq. (\ref{omori-expo}).
For the $q$-dependent relaxation (Fig. \ref{f5}), the results obtained using Eq. (\ref{max}) reveal some differences with the analytical estimate (\ref{omori-q2}). On one hand, we see the appearance of a small time cut off in the power law decay. This cutoff is due to a large $q$ (or small distance) cutoff in the model implied by the finite lattice parameter.
Although this effect is not contained in the analytical estimation Eq. (\ref{omori-q2}), it is easily obtained considering a maximum $q$
when integrating Eq. (\ref{cutoff}). The qualitative form of the result reproduces very well the effect of the $c$ parameter in the Omori
expression, Eq.  (\ref{omori}).
In addition, the decay exponent predicted by Eq. (\ref{omori-q2}) is a bit smaller than the actual result for $N(t)$ calculated using Eq. (\ref{max}). The discrepancy is actually not too large (amounting approximately to a difference of 0.15 in the value of the exponent), and is rather independent of the roughness of the initial surface. \cite{footnote3}

\begin{figure}[h]
\includegraphics[width=8cm,clip=true]{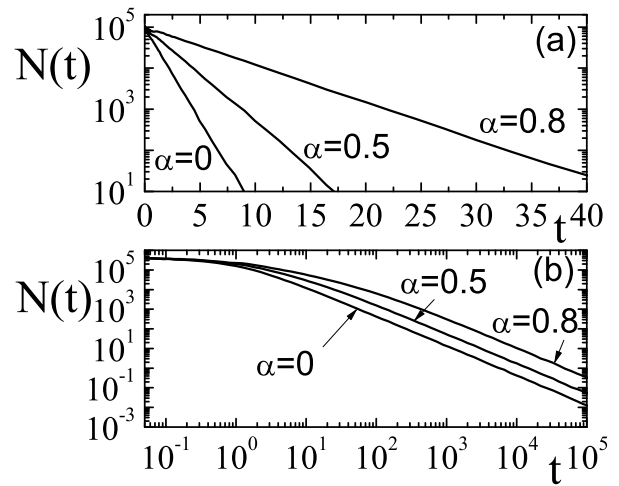}
\caption{Effect of aftershocks triggering other aftershocks. Primary aftershocks are generated with a rate $N(t)$. In (a) $N(t)\simeq \exp (-t)$, in (b) $N(t)\simeq 1/(t+c)^p$ (we take $p=1.5$, $c=1$, for concreteness).  Each aftershock can trigger successive ones with probability $\alpha$. We see that the effect of a finite $\alpha$ is to change the time decay constant in the exponential case (a), whereas in the Omori case (b) it has a minor effect on the form of the decay rate.
}
\label{otra}
\end{figure}

The aftershock rate as given by Eq. (\ref{max}), can be named that of ``primary" aftershocks.
A more accurate estimation of the aftershock rate must account for the modifications in the values of $f_i$ that every triggered aftershock generates.
A qualitative analysis of the kind of effect we must expect is the following. If $N(t)$ is the rate of primary aftershocks and if a primary aftershock actually occurred at time $t_0$, it refreshes the values of $f_i$ in the region it affected, resetting the production of aftershocks in this region, 
that becomes now proportional to $N(t-t_0)$. A full counting of all aftershocks must take this effect into account. The full effect is difficult to assess analytically, but to make some quantitative estimation, we proceed as follows. We consider an initial aftershock from a primary distribution $N(t)$. Let us assume it occurred at time $t_0$. Now, it is supposed that this primary aftershock has a probability $\alpha$ of generating a secondary one, at a time
defined by the distribution rate $N(t-t_0)$. If it actually occurs, it can generate a third one, etc. In this way, with the primary rate $N(t)$ and some assumed value of $\alpha$ we can have an indication of the full aftershock rate. In Fig. \ref{otra} we see the effect of finite $\alpha$ values on two typical forms of $N(t)$, namely the exponential form $N(t)\sim \exp (-t)$, appropriate for model A, and the Omori form  $N(t)\sim 1/(t+c)^p$, appropriate for model B.

In the exponential case, we see how increasing values of $\alpha$ change the time constant decay of the exponential distribution to larger values.
For the Omori case, there is a visible effect that can be described as an effective increase of the $c$ value, however, for large times the value of the power law exponent is not modified.

\begin{figure}[h]
\includegraphics[width=8cm,clip=true]{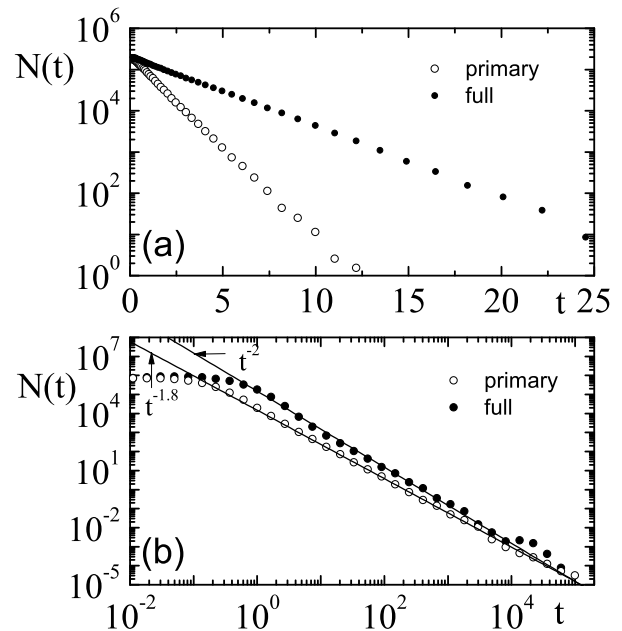}
\caption{Aftershock rates obtained using a starting configuration corresponding to an equilibrium qEW surface. Open symbols correspond to the case in which each triggered aftershock is not allowed to modify the distribution of $f_i$. Full symbols instead, correspond to a case in which each aftershock modifies the values of $f_i$ according to the avalanche dynamics (Eq. \ref{avalanche1}). Panels (a) and (b) correspond to models A and B (Eqs. (\ref{d+r_1}) and (\ref{d+r_2})).
}
\label{f6}
\end{figure}

Now we turn to numerical simulations in the actual viscoelastic models to try to confirm this behavior.
We proceed as follows. An initial surface from the purely elastic qEW model is generated as before, keeping track of the values of forces $f_i$, interface positions $x_i$, and the thresholds forces $f_i^{th}$. We make two different evolution algorithms. In one of them $f_i$ is relaxed in time, and each time $f_i+k_0(Vt-x_i)$ reaches $f_i^{th}$ we count one aftershock, but the values of $f_i$ and the position of the interface are not modified. This is actually quite similar to the analysis in Figs. \ref{f4} and \ref{f5}, and is made only for comparison purposes. The aftershocks counted in this way correspond to what we have called the primary aftershocks.
In the second algorithm, each time $f_i+k_0(Vt-x_i)$ reaches $f_i^{th}$, we fully develop the avalanche, modifying the form of $f_i$ according to the evolution equations (\ref{avalanche1}). This changes the occurrence of ulterior aftershocks, and modifies the aftershock rate. We refer to this as the {\em full counting} of aftershocks.
The comparison between the two situations is presented in Fig. \ref{f6}.
It displays qualitatively the effect presented in Fig. \ref{otra}. In the case of model A, full counting produces a decrease in the time decay constant of the distribution, which however continues to be roughly exponential. In the case of model B, the full counting rate of aftershocks continues to be a power law decay. The decay exponent seems to be in this case somewhat larger than the one obtained considering only primary aftershocks.

\begin{figure}[h]
\includegraphics[width=8cm,clip=true]{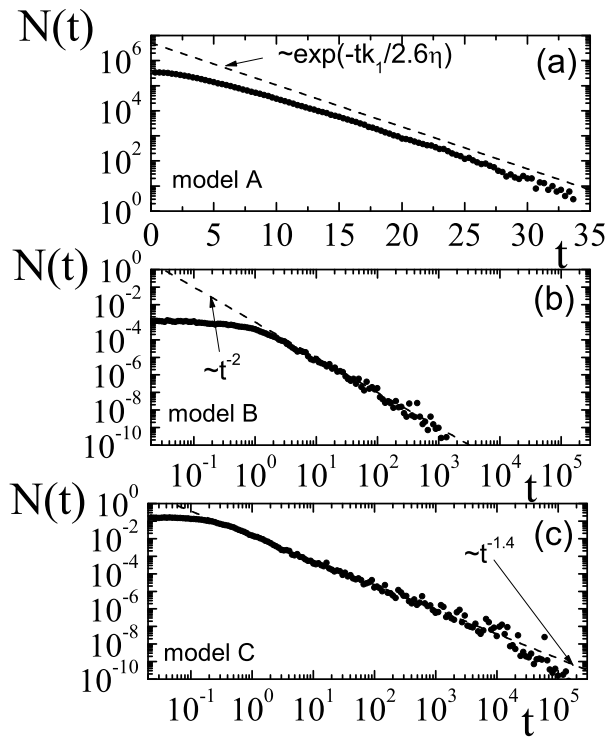}
\caption{Aftershock rates in full simulations of the viscoelastic model studied in this paper. (a) In the single relaxation time case (model A, Eq. (\ref{d+r_1})) the aftershock rate is roughly exponential, with a decay time larger than the relaxation time in the system. (b) 
For $q$-dependent relaxation (model B, Eq. (\ref{d+r_2})), an Omori law is obtained, with a value of the decay exponent close to 2. 
(c) Aftershock rate for model C (Eq. \ref{modelc}).
}
\label{f7}
\end{figure}

Finally, we now present the  results obtained from full simulations of the viscoelastic models A and B. Namely, we let the system evolve for a long time until it reaches a stationary state, and follow and record the activity in the system in this regime (obtaining sequences as the one presented in Fig. \ref{f3}). 
We classify all aftershocks by its occurrence time with respect to the initial event in the cluster. By definition, these are strictly positive values. We then make the statistics of these times. The results are contained in Fig. \ref{f7}.
For model A the aftershocks show an exponential decay (with some deviations at short times) with an effective time constant that is larger (in a factor $\sim 2.6$ in this case) than the bare value $\eta/k_1$.
For model B, an Omori law, with a short time cutoff and a power law distribution at long times is clearly observed. The value of the decay exponent is close to 2, compatible with the previous results (Fig. \ref{f6}) corresponding to an interface with a roughness similar to the original qEW model.
This is an indication  that the distribution of $f_i$ in the full viscoelastic model after big shocks has a similar roughness as in the original qEW model. This is not surprising: since avalanches occur instantaneously, they are unaffected by the viscous elements in the model, namely they occur exactly as in a qEW model.

\section{Summary and Discussions}

In this paper we have studied the statistics of aftershocks produced in interface depinning models that include viscoelastic relaxation. Two cases were analyzed in detail, namely the single relaxation time case (model A, Eq. \ref{d+r_1}), and a case of $q$-dependent relaxation (model B, Eq. \ref{d+r_2}). These two cases can be conveniently described by simple mechanical analogs, using springs and dashpots (Fig. \ref{f1}). The model with a single relaxation time produces aftershocks decaying in time exponentially. The observed time constant of the aftershock rate is larger than the system relaxation time due to the cumulative effect of secondary aftershocks (those triggered by previous aftershocks). For the $q$-dependent relaxation an analytical estimate gives a power law decay of the aftershock rate, with a numerical value of the exponent $p=(4-\zeta)/2$, with $\zeta$ being the roughness exponent of the avalanches that occur in the system.
Numerical simulations confirm the power law decay form, although they yield numerical values of the exponent somewhat larger than the analytical prediction. In particular, we measure $p\simeq 2$. The effect of secondary aftershocks is much less noticeable in this case.

One of our main conclusions of this paper is that the $q$-dependent relaxation that appears for certain types of viscoelastic relaxation generates a power law decay of the aftershock rate, compatible with the phenomenological Omori law for earthquakes. 
Yet, the measured value $p\simeq 2$  is higher than typically reported values in actual seismicity (between 0.9 and 1.5 \cite{utsu2}).
We want to discuss on possible mechanisms that may produce smaller, more realistic values of $p$ in the context of  viscoelastic models.
 
One possibility is to have a different $q$-dependence of the relaxation mechanism. We have discussed in detail the case $\dot f\sim - f$ producing an exponential decay of aftershocks, and $\dot f\sim \nabla ^2 f$ that gives a potential decay, with an analytically estimated $p=(4-\zeta)/2$. 
A higher order relaxation mechanism, as for instance

\begin{equation}
\frac {df}{dt}\sim -\nabla ^4 f~~~~~~\mbox{(model C)}
\label{modelc}
\end{equation}
(which has in fact been shown to be plausible for some viscoelastic models \cite{kolton}), gives a smaller $p$, in particular for this case it is analytically estimated (on the same lines as before) that  $p=(6-\zeta)/4$. 
The result of numerical simulations with this model is shown in Fig. \ref{f7}(c) and show that in fact the aftershock rate of this model is power law, with $p\simeq 1.4$. This makes clear that different relaxation mechanism produce different aftershock rates, with higher order mechanisms giving rise to smaller $p$ vales.

A second mechanism giving rise to a smaller $p$ value is realized in the model studied in Ref. \cite{jagla}. There, the viscoelastic relaxation is supposed to affect not only the force exerted by the $k_1$ springs (see Fig. \ref{f1} here) but also that exerted by the driving springs $k_0$. 
This implies that the forces that are relaxed have a contribution $f_i\sim k_0 x_i$ [in addition to the component $f_i\sim k_1(\nabla^2 x)_i$, (Eq. \ref{fi})]. An analytical estimation of the consequence of this fact, leads to a much smaller value of $p$, as in fact it is effectively observed in actual simulations with these kind of models (see \cite{jagla}, Figs. 6 and 9).

The dependence of the aftershock rate on the precise relaxation mechanism is interesting in view of the actual variation of $p$ observed in different geographical locations\cite{utsu2}. It may be expected that a deeper understanding of the relation between the $p$ value and the kind of relaxation mechanism at play can give information on what is the relevant relaxation mechanism in different geographical locations. 

We want to finish with a qualitative discussion concerning the nature of the mechanism producing aftershocks in the present viscoelastic models, compared with the mechanism originally proposed in \cite{dietrich1994}. 
The traditional mechanism assumes that aftershocks occur in regions that are overloaded due to the stress redistribution caused by the initial shock. This implies at once that aftershocks are not expected inside the rupture region of the initial shock, where stress has decreased. This is at odds with the observation that a majority of aftershocks occur within the initial rupture region. Helmstetter and Shaw \cite{helmstetter} have shown how this can be explained assuming a stochastic model of the stress redistribution within the initial rupture region. Combined with an assumed rate-and-state friction law, they obtain a realistic Omori law for the aftershock rates. The models we study here are, on one hand, inherently stochastic as the position $x_i$ and strength $f^{th}$ of the pining centers are stochastic variables. The existence of aftershocks depends crucially on this assumption, and the aftershocks always occur within the region affected by previous shocks in the same cluster. On the other hand, we do not need to assume the validity of the rate-and-state description of the sliding process. A phenomenology compatible with rate-and-state friction emerges naturally from the microscopic relaxation mechanisms introduced \cite{kolton}. Yet, an unrealistic feature of our models is the consideration of only local elastic interactions, whereas it is well known that due to the three-dimensional nature of the full problem long range elastic interaction should be considered. We expect that realistic long range interaction will have an effect in the aftershock rate, as in particular, long range elastic interactions modify the typical roughness of qEW interfaces. 
Also, in the presence of long range interactions, a fraction of the aftershocks can nucleate outside the region affected by previous shocks, giving rise to a more realistic situation.
Unfortunately, the precise assessment of the effects caused by the consideration of long range elastic interactions in models with viscoelastic relaxation can only be addressed by costly numerical simulations that are out of our present possibilities.

\section{Appendix: Numerical Details} 

The narrow pinning centers are chosen to be randomly distributed along the $x$ axis, with an average separation $\overline z=0.1$. The results are independent of this particular choice.
Pinning centers  are uncorrelated among different spatial positions. For the numerical implementation, each time the interface moves forward, the location of the new narrow well is obtained by adding to the previous position a quantity $z$, that is exponentially distributed, with mean value $\overline z$. This generates a random uncorrelated distribution for the location of the wells.

The value of the threshold forces $f^{th}$ at each pinning center is taken from a Gaussian distribution with mean value 1 and standard deviation 1. Negative values are discarded. As these values are uncorrelated for different pinning centers, each time the interface jumps to a new position, the value of $f^{th}$ is chosen anew.

The size $S$ of an avalanche is defined as the sum of all displacements at every point on the interface, namely $S=\sum_i \delta x_i =\sum_i (x_i^{\text {after}}-x_i^{\text {before}})$, where before and after refer to the values of $x_i$ before the beginning and after the end of the avalanche. 

Throughout the paper, the value of $k_1$ is set to $k_1=1$. The value of $k_0$ is $0.05$ in Fig. \ref{f3} and in the construction of the qEW surface in Figs. \ref{f4} and \ref{f5}, and 0.15 in Figs. \ref{f6} and \ref{f7}.
The spatial numerical lattice is an  $N\times N$ square, with periodic boundary conditions. The value of $N$ is 1024 in all cases except in Fig. \ref{f7}(b-c), where it is 256.

\section{Acknowledgments} 

I thank Alberto Rosso and Fran\c cois Landes for stimulating discussions and for the suggestion of model B as a case worth of consideration.
This research was financially supported by Consejo Nacional de Investigaciones Cient\'{\i}ficas y T\'ecnicas (CONICET), Argentina. Partial support from
grant PICT-2012-3032 (ANPCyT, Argentina) is also acknowledged.


\begin{thebibliography}{4}
\bibitem{footnote} Note that in the two viscoelastic models studied here, if $x_i$ are constant, $f_i$ tend to zero with time, meaning that all the elastic energy of the interface is viscoelastically relaxed. It would be straightforward to consider a more realistic case in which only a fraction of the elastic energy of the interface is relaxed (see Ref. \cite{landes}). This however complicates the presentation, without adding to the physics of the problem.
\bibitem{footnote3}Further analysis shows that Eq. (\ref{omori-q2}) actually fits very well the quantity given by Eq. (\ref{ndet2}).
This means that the slight difference is due to the non-perfect equivalence of expressions (\ref{max}) and (\ref{ndet2}).
\end{thebibliography}
\end{document}